\documentclass[11pt]{article}
\usepackage[cp1251]{inputenc}
\usepackage[T1]{fontenc}
\usepackage{textcomp}
\usepackage[centertags]{amsmath}
\usepackage{amsfonts}
\usepackage{amssymb}
\usepackage{revsymb}
\usepackage[pdftex]{hyperref}
\usepackage{graphicx}
\usepackage{graphbox}
\usepackage[numbers,sort&compress]{natbib}

\usepackage{paperinitial}

\paperinitialization{15mm}{15mm}{15mm}{15mm}{2pt}{10pt}

\newcommand{\e}{\varepsilon}
\newcommand{\vf}{\varphi}

\newcommand{\s}{\sigma}

\newcommand{\be}{\beta}
\newcommand{\ga}{\gamma}

\newcommand{\spx}{\mathbf{x}}

\begin{document}
\allowdisplaybreaks[4]
\frenchspacing
\setlength{\unitlength}{1pt}

\title{{\Large\textbf{Proposal for the experimental observation of twisted photons in transition and Vavilov-Cherenkov radiations}}}

\date{}

\author{O.V. Bogdanov${}^{1),2)}$\thanks{E-mail: \texttt{bov@tpu.ru}},\; P.O. Kazinski${}^{1)}$\thanks{E-mail: \texttt{kpo@phys.tsu.ru}},\; and G.Yu. Lazarenko${}^{1),2)}$\thanks{E-mail: \texttt{laz@phys.tsu.ru}}\\[0.5em]
{\normalsize ${}^{1)}$ Physics Faculty, Tomsk State University, Tomsk 634050, Russia}\\
{\normalsize ${}^{2)}$ Division for Mathematics and Computer Sciences},\\
{\normalsize Tomsk Polytechnic University, Tomsk 634050, Russia}}

\maketitle

\begin{abstract}

We propose to use Vavilov-Cherenkov (VC) and transition radiations as a source of twisted photons in a wide range of energies. The experimental setup to observe the orbital angular momentum of photons constituting those radiations is described. The radiation produced by relativistic electrons and ions in passing dielectric plates is considered. The experimental verification of the addition rule for the total angular momentum in the radiation of twisted photons by helically microbunched beams is proposed for VC and transition radiations. The parameters of the experiments are presented.

\end{abstract}

\section{Introduction}

The transition and Vavilov-Cherenkov (VC) radiations of plane-wave photons are well studied and find various applications \cite{Ginzburg, Pbook, x-ray, Karataev, ArutOgan94}. These radiations allow one to generate the photons from the THz spectral range \cite{ACurcio19} to X-rays \cite{x-ray, x-ray VCh1, x-ray VCh2}. The properties of VC and transition radiations are employed for elaboration of detectors of charged particles \cite{x-ray VCh1, x-ray VCh2, TR det}. The analysis of these radiations is used for diagnostics of the form of particle bunches and the energy distribution of particles in them \cite{x-ray, Karataev, x-ray VCh1, x-ray VCh2, Tishchenko}.

As was shown in \cite{OAMeVCh, OAMeTR1, OAMeTR2, Kaminer, KnyzSerb, BKL5, BKLb, ExpHemsing, TrHemsing}, the radiation by relativistic charges moving in or near the medium possesses new properties due to the orbital angular momentum of radiating or radiated particles. Usually, the conversion of plane-wave photons to the twisted ones is achieved by the use of phase or holographic plates \cite{KnyzSerb, xrayplate}. This approach cannot be employed for development of the brilliant sources of hard twisted photons. The use of transition and VC radiations allows one to generate twisted photons in a wide spectral range up to X-rays \cite{BKL3, BKL5, BKLb}. The helically microbunched beams of charged particles \cite{HemStuXiZh14} propagating in a medium produce the coherent transition and VC radiations with large intensity and angular momentum \cite{BKLb, BKL5, HKDXMHR}, what is not achievable by using the phase and holographic plates at the present moment.

In comparison with the plane-wave photons, the twisted photons possess an additional degree of freedom -- the projection of the total angular momentum $m$ on the propagation axis. This additional degree of freedom was used for overcoming the diffraction limit \cite{diflim}, to increase the capacity of telecommunication channels \cite{2^m, key}, and for quantum cryptography \cite{key}. One of the property of twisted light is that its intensity vanishes at the center of the beam. This allows one to design the optical tweezers and to manipulate nanoparticles \cite{trapp,Greir}. The distribution of radiation over $m$ is rather sensitive to the structure of the beam of radiating particles. This property can be employed for an improvement of the VC detectors \cite{Kaminer} and the detectors based on the transition radiation \cite{ExpHemsing, TrHemsing}.

\section{Theory}

The twisted photons are the quanta of the electromagnetic field with the definite energy $k_0$, the momentum projection $k_3$, the projection of the total angular momentum $m$ on the same axis, and the helicity $s$. We suppose that the axis $3$ coincides with the detector axis. The absolute value of the momentum component perpendicular to the detector axis is $k_\bot =\sqrt{k_{0}^{2}-k_{3}^{2}}$.

\begin{figure}[tp]
\centering
\includegraphics*[width=0.8\linewidth]{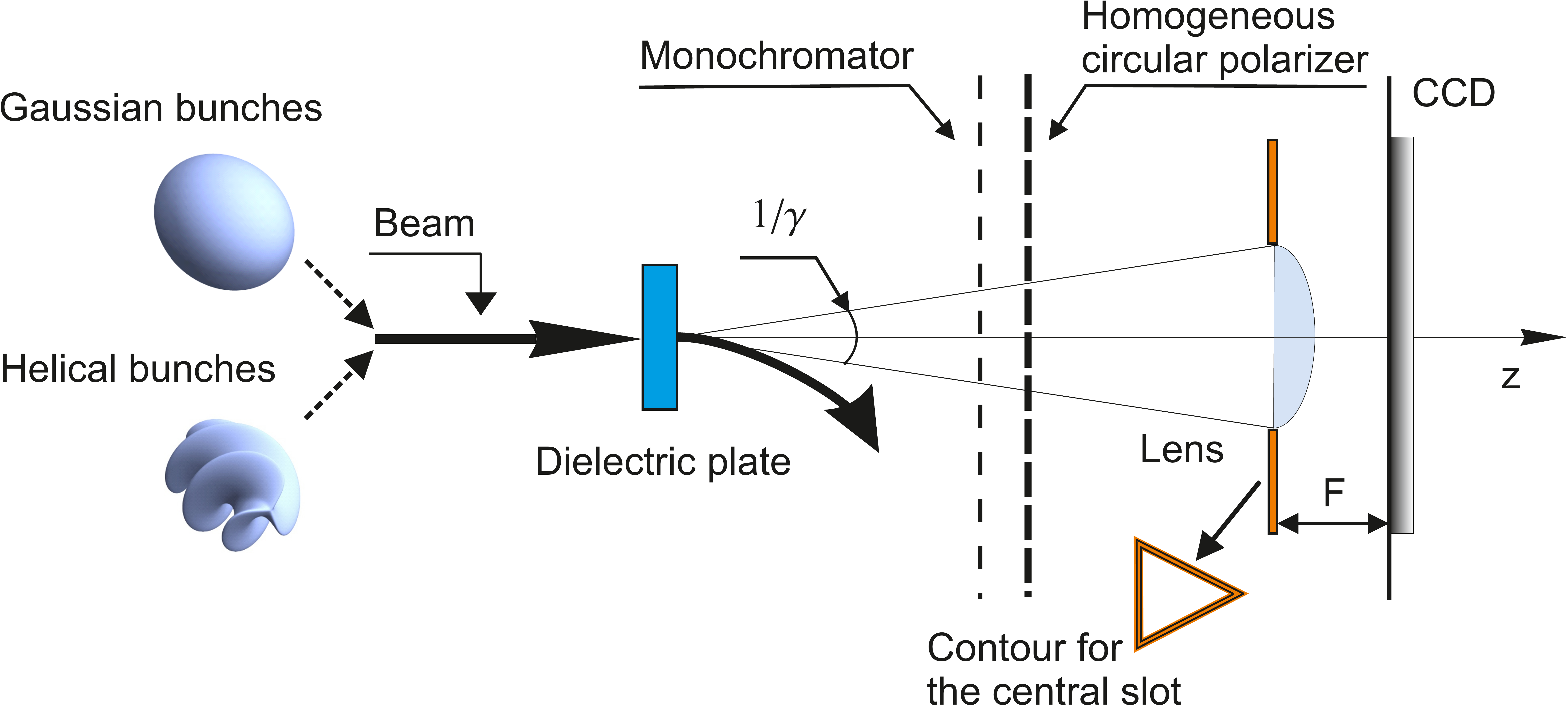}
\caption{{\footnotesize The scheme for the experimental setup.}}
\label{Experiment design}
\end{figure}

In the paper \cite{BKL5}, the general theory of radiation of twisted photons in the presence of a dispersive medium was developed. The probability of radiation of a twisted photon by a charged particle takes the form
\begin{equation}\label{prob_plate}
    dP_1(s,m,k_\perp,k_3)=e^2|a|^2\bigg|\int_{-\infty}^\infty d\tau e^{-ik_0 x^0 (\tau)}
    \big(\dot{\spx}(\tau),\boldsymbol{\Phi}(s,m,k_\perp,k_3)\big)\bigg|^2 \Big(\frac{k_\perp}{2k_0}\Big)^3\frac{dk_3 dk_\perp}{2\pi^2},
\end{equation}
where $x^\mu(\tau)$ is the particle's worldline, $|a|$ is the normalization constant, and $\boldsymbol{\Phi}(s,m,k_\perp,k_3)$ are the mode functions of twisted photons (see for details \cite{BKL5}). We consider the case when the ordinary Gaussian and helically microbunched beam of identical particles falls normally onto the dielectric plate. In that case, the probability of radiation of twisted photons can be found from the formula
\begin{equation}\label{prob_rad_tot_hw}
    dP_\rho(s,m,k_\perp,k_3)=\big[Nf_{m}+N(N-1)|\vf_{m}|^2\big]dP_1(s,0,k_\perp,k_3),
\end{equation}
where $dP_1(s,0,k_\perp,k_3)$ is the probability of radiation of a twisted photon by one charged particle moving along the center of the beam. The functions $f_{m}$ and $\vf_m$ are the incoherent and coherent interference factors, respectively (see for details \cite{BKb,BKLb, BKL5}).

\section{Experiment and results}

The scheme of the experiment is presented in Fig. \ref{Experiment design}. The beam of relativistic charged particles moves along the detector axis (the axis $z$ or $3$), traverses the dielectric plate, and then is deflected. The parameters of the beam and the dielectric plate are chosen such that the created transition and VC radiations are concentrated in a narrow cone. As for transition radiation, the maximum of its intensity is located at $n_\perp:=k_\perp/k_0=\sqrt{3}/\gamma$. The VC radiation is concentrated at $n_\perp=\sqrt{\e'(k_0)-\be^{-2}}$, where $\e'(k_0)$ is the real part of the permittivity. We have to demand $n_\perp\ll1$, i.e., the paraxial approximation for radiation must be applicable. In this approximation, $m=l+s$, where $l$ is the orbital angular momentum of a twisted photon. The created transition and VC radiations pass sequentially through the monochromator separating a narrow spectral band and the homogeneous circular polarizer distinguishing the helicity $s=1$ or $s=-1$. In fact, the monochromator is only needed for the successive circular polarizer as, usually, the latter works only in the narrow spectral band. Then the radiation falls onto the detector of twisted photons. The simplest detector of pure states of twisted photons is the triangle aperture \cite{aperture, taira2019, GoTsuKu} with the lens. The diffraction pattern formed by this aperture and recorded by the CCD camera, which is located at the focal distance $F$ from the lens, has a rather peculiar form for the infalling radiation with definite orbital angular momentum. This diffraction pattern represents a collection of spots \cite{aperture, taira2019, GoTsuKu} and the number of spots arranged along the perimeter of the triangle is equal to $3|l|$. The size of the aperture should be slightly smaller than the size of the spot of the infalling radiation. Such a detector of twisted photons operates in different spectral ranges. The experiments \cite{taira2019, GoTsuKu} show that the lens in this scheme can be removed without a large distortion of the diffraction pattern.

\begin{figure}[tp]
\centering
\includegraphics*[width=0.48\linewidth]{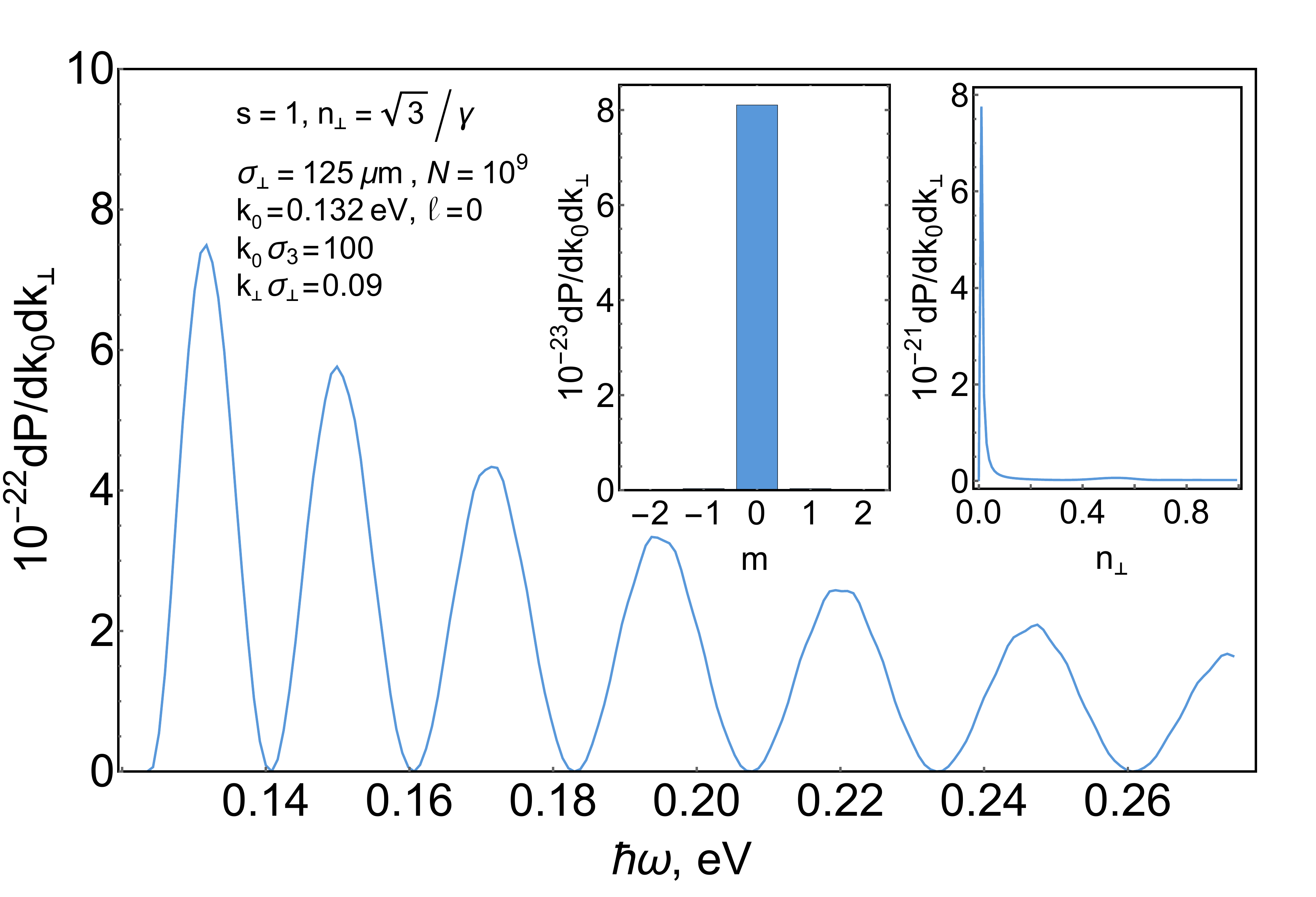}\;
\includegraphics*[width=0.475\linewidth]{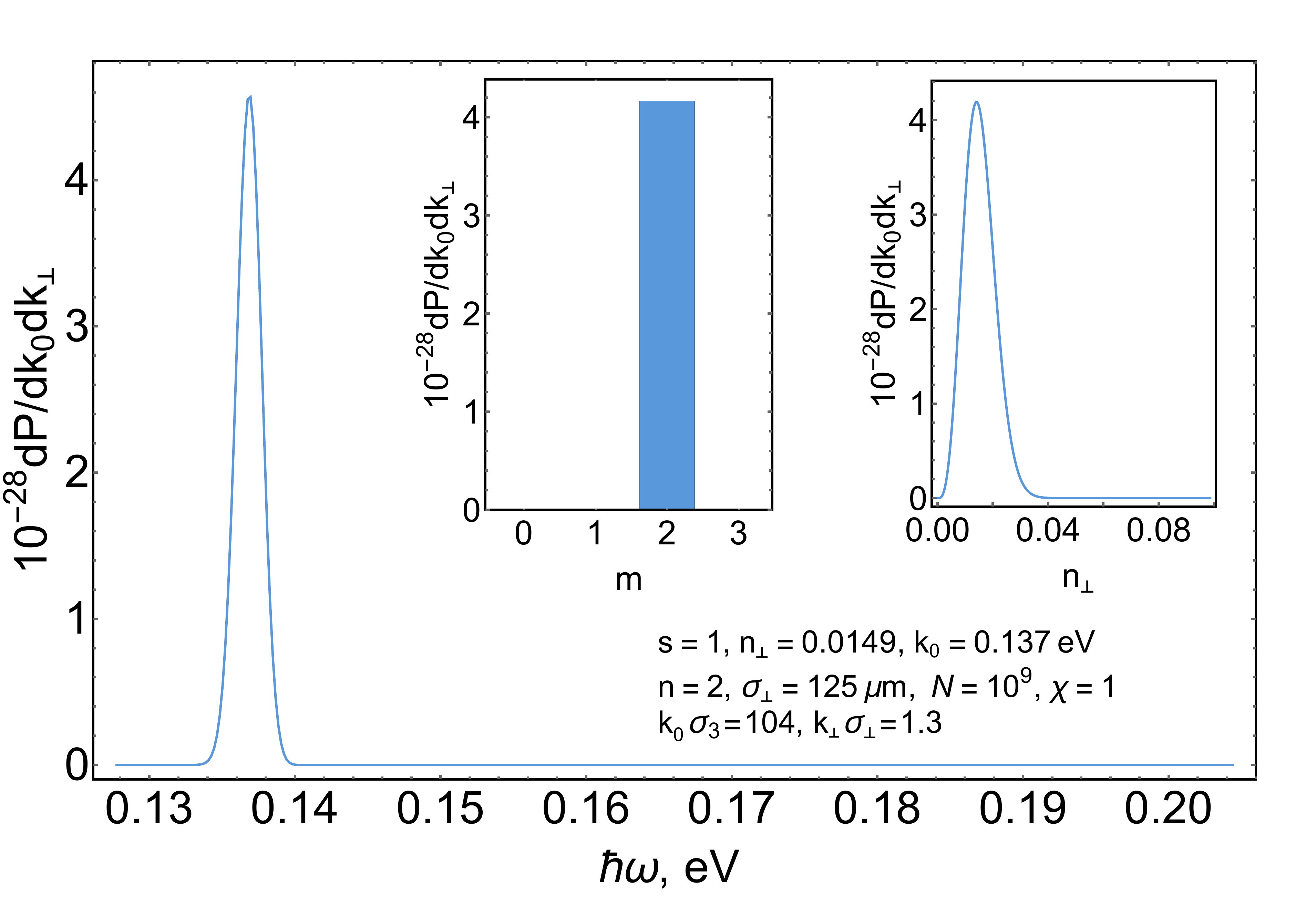}
\caption{{\footnotesize The transition radiation from charged particles traversing the LiF dielectric plate of the thickness $L=100$ $\mu$m. The electrons move along the detector axis towards the detector. The number of electrons in the beam is $N=10^9$ and their Lorentz factor is $\ga=1673$, $E=855$ MeV. The beam is supposed to have a Gaussian profile with the longitudinal dimension $\s_3=150$ $\mu$m (duration $0.5$ ps) and the transverse size $\s_\perp=125$ $\mu$m. Left panel: The transition radiation of the uniform Gaussian beam of particles. The contribution of coherent radiation is strongly suppressed. The total angular momentum per photon $\ell=0$. The small hump at $n_\perp\approx0.6$ is the VC radiation. Right panel: The transition radiation of the helically microbunched beam of particles. The fulfillment of the addition rule on the second coherent harmonic of radiation is clearly seen. The total angular momentum per photon $\ell=2$.}}
\label{dielectricplate}
\end{figure}

\begin{figure}[tp]
\centering
\includegraphics*[width=0.48\linewidth]{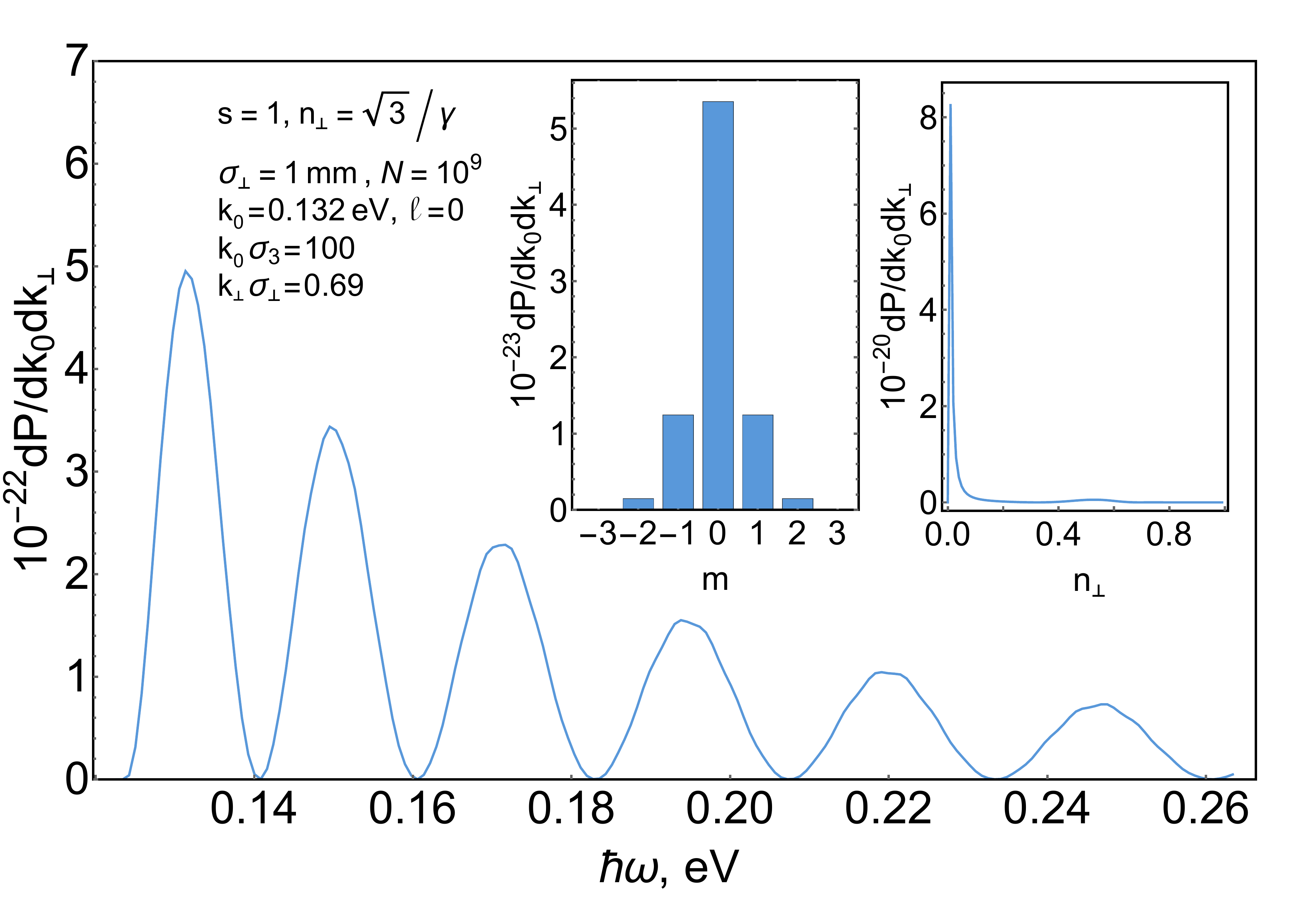}\;
\includegraphics*[width=0.49\linewidth]{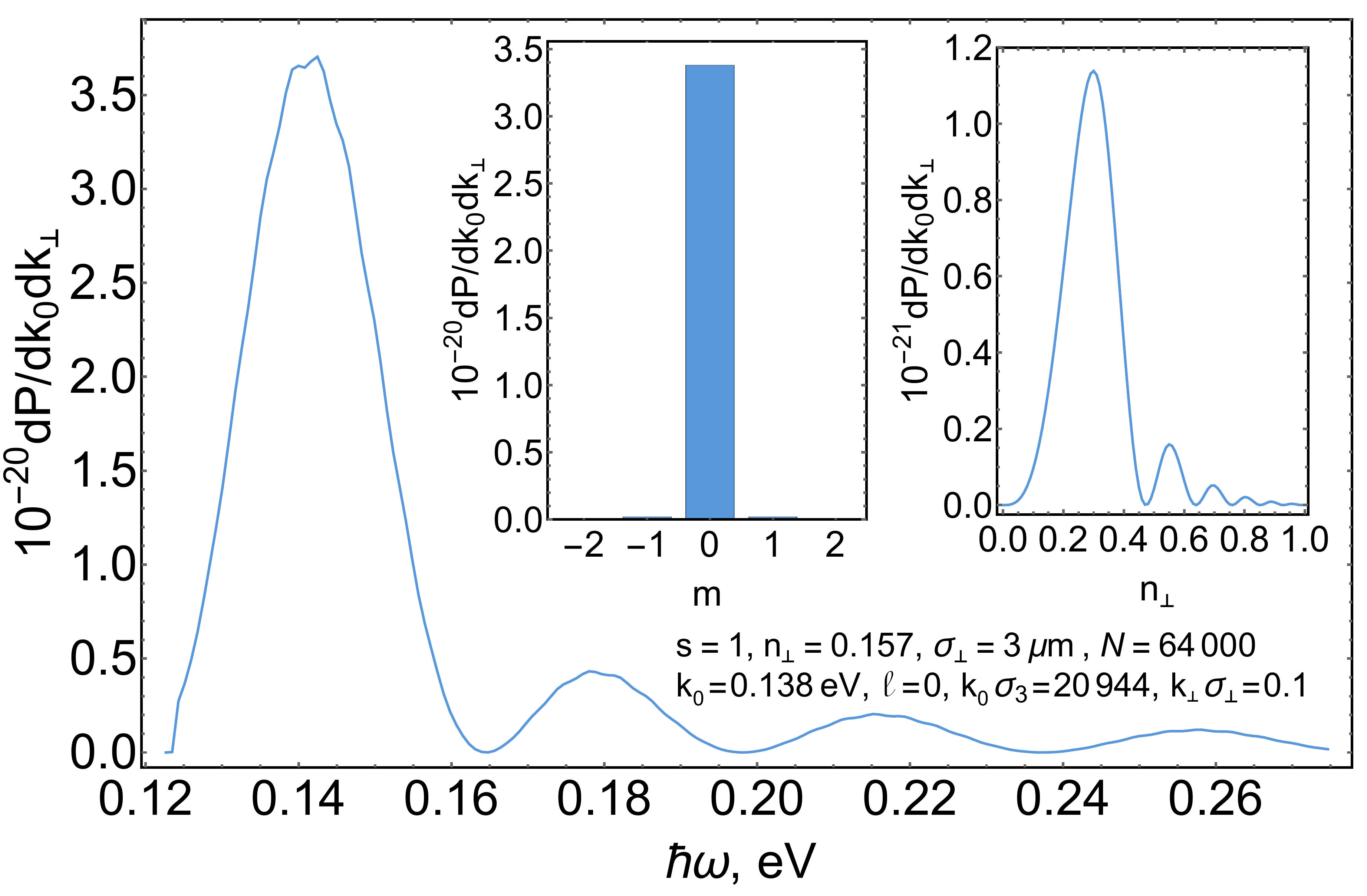}
\caption{{\footnotesize Left panel: The same as on the left panel in Fig. \ref{dielectricplate} but for $\s_\perp=1$ mm. The distribution of radiated twisted photons with respect to $m$ spreads over a wider interval. The total angular momentum per photon $\ell=0$. Right panel: The VC radiation from the bare ions of ${}^{238}_{92}$U traversing the LiF dielectric plate of the thickness $L=100$ $\mu$m with the Lorentz factor $\ga=2$, the kinetic energy $T=1$ GeV per nucleon. The bunch is supposed to have a Gaussian profile with the longitudinal dimension $\s_3=3$ cm (the duration $100$ ns) and the transverse size $\s_\perp=3$ $\mu$m. The number of particles in the bunch $N=6.4\times 10^3$. The VC radiation is concentrated at the angle $n_\perp=0.157$. The total angular momentum per photon $\ell=0$.}}
\label{condplateandVChion}
\end{figure}

As follows from \eqref{prob_rad_tot_hw} and the properties of the interference factors, the pure source of twisted photons is obtained when a narrow beam of charged particles radiates. It should be such that $k_\perp\sigma_\perp\ll1$, where $\s_\perp$ is the typical transverse size of the beam. When this condition is satisfied, the ordinary (uniform) Gaussian beam of relativistic charged particles traversing the dielectric plate radiates the twisted photons with $m=0$ (cf. the plots in Figs. \ref{dielectricplate} and  \ref{condplateandVChion}). Therefore, in the paraxial regime, the orbital angular momentum of this radiation is $l=-s$. The use of helically microbunched beams of particles instead of the ordinary Gaussian ones allows one to shift the total angular momentum of radiated twisted photons. Namely, for the configuration considered, the total angular momentum of twisted photons radiated at the $n_c$-th coherent harmonic is $\chi n_c$, where $\chi$ is the chirality of the helical beam (see for details \cite{BKLb,BKL5}). The shift of the orbital angular momentum in the radiation produced by the helically microbunched electron beam evolving in the planar undulator was observed experimentally in \cite{HKDXMHR}. The technique for production of helical beams is described in \cite{HemStuXiZh14}.

The use of ion beams in production of twisted photons by means of transition and VC radiations is more advantaged in comparison with the electron beams so long as the intensity of radiation of ion beams is $Z^{2}$ times more intense, where $Z$ is the charge number (see Fig. \ref{condplateandVChion}). The moderate Lorentz factors of ion beams lead to a small VC angle of radiation and so the paraxial approximation holds. Unfortunately, the design of helically microbunched beams of heavy ions is an open problem at present.

\section{Conclusion}

The prospective results of the proposed experiments:
\begin{enumerate}
  \item It will be shown for the first time that the transition and VC radiations possess the orbital angular momentum $l=-s$ for narrow relativistic Gaussian charged particle beams.
  \item It will be experimentally verified for the first time that the addition rule $m=\chi n_c$ holds in the transition and VC radiations produced by narrow helically microbunched beams of charged relativistic particles.
\end{enumerate}

As far as transition radiation is concerned, the same experiments can be conducted with the metallic plate instead of the dielectric one.

\paragraph{Acknowledgments.}

We are grateful to P. Karataev and D. Karlovets for the fruitful conversations during the symposium RREPS-2019. The work was supported by the Russian Science Foundation (project No. 17-72-20013).

\end{document}